# Quantifying the efficacy of childcare services on women employment

Jing-Yi Liao[1], Ying Kong[1,*], Tao Zhou[2,*]

1 Shenzhen International Graduate School, Tsinghua University, Shenzhen 518055, People's Republic of China
2 CompleX Lab, University of Electronic Science and Technology of China, Chengdu 611731, People's Republic of China
* Corresponding Authors: kongying@sz.tsinghua.edu.cn (YK)
zhutou@ustc.edu (TZ)

## Abstract

Women are set back in the labor market after becoming mother. Intuitively, childcare services are able to promote women employment as they may reconciliate the motherhood penalty. However, most known studies concentrated on the effects of childcare services on fertility rate, instead of quantitative analyses about the effects on women employment. Using worldwide panel data and Chinese data at province level, this paper unfolds the quantitative relationship between childcare services and women employment, that is, the attendance rate of childcare services is positively correlated with the relative employment rate of women to men. Further analysis suggests that such a positive impact may largely resulted from breaking the vulnerable employment dilemma.

# Introduction

According to the data from International Labor Organization, on average, women's participation in the labor market is lower than men over the world, and the gap has barely narrowed in the past 30 years. Many women have been forced to leave the workplace due to becoming mother. Facing substantial disadvantages, mothers are judged as being less dependable, authoritative and committed to their jobs but more emotional and irrational than non-pregnant women [1]. But even for these women without child, their employment chances are affected, because of the pregnancy risks during their childbearing age and the unpleasure caused by the violation of prescribed social role if they decide not to have child [2]. This phenomenon is depicted by a well-known concept *motherhood penalty*, which refers to the negative impacts of the birth event on women career development, mainly resulted from the contradiction between paid and unpaid job. It takes effect from fertilization to postpartum, and lasts in the whole process of raising a child [3]. Negative effects due to motherhood penalty include the limitations of career choice, hiring, promotion, wage gap, and so on [4]. Although income is the most common indicator about motherhood penalty in recent studies, being able to participate in the workplace is the premise of earning. Therefore, this article focuses on the extent of women's participation in the workplace.

There are three different yet related mechanisms to explain the negative effects caused by motherhood penalty [3]. *Family restriction theory* addresses that once women become mother, they tend to further devote their time biased towards childrearing and other unpaid household labors [5,6], otherwise society would recognize them as unqualified mothers. Therefore, their availability in the paid work is reduced. *Human capital theory* believes that the interruption from the birth event will eliminate the value of the workforce [7]. Even mothers are not completely leaving their jobs, they might choose family-friendly positions instead [4], which provide convenient to taking care of children but trade off their career prospects [8]. *Employer discrimination theory* is the combination of preference discrimination and statistical discrimination [9]. While women are expected to be good mothers and housewives, those who enter the labor market and compete with men are usually unwelcome [2], causing negative impacts on career development [10]. Apart from the preference discrimination, with insufficient information, employers tend to evaluate individuals by using statistics on the group they belong to, in which women are regarded as less productive resulted from historical preference discrimination. In all, the major link to motherhood penalty lies in the opportunity cost of raising a child.

The childcare policies designed to transferring and sharing the cost of childrearing among government, labor market, family and women themselves would affect the extent of motherhood penalty. These instruments are divided into three categories: childcare leave, childcare subsidy and childcare services. While the extended parental leave could reduce the conflict between work and family, it somehow suggests to strengthen women's role as mother [11]. Moreover, the financial incentives on childcare seems to reduce the employment rate of household second earners, which normally refer to women [12]. Although different policies interact with each other, in general, the efficacy of childcare services are stronger than other measures in supporting women in the workforce [13]. Based on all above, this article hypothesizes that the provision of childcare service is highly effective to promote labor force participation rate of women.

This article only accounts for center-based childcare services. The reasons are twofold. Firstly, public policies and subsidies from government can only affect center-based ones, rather than other types of childcare services like those by childminders [14]. Secondly, although there are various sources of low-cost or even free childcare from spouses, grandparents, teenagers, other relatives and friends, they are influenced by many observable or hidden societal, economic and cultural factors that are not easy to be quantified. Most known studies concentrated on the effects of childcare services on fertility rate, while quantitative analyses about the relationship between childcare services and female labor supply are insufficient and have not drawn a consistent conclusion. It is widely believed that childcare services play as the liberator to enable women back to work [15], while Thévenon recently argues that precondition is needed for childcare services to play an active role, for example, childcare services play a significant role only for countries with relatively high degrees of employment protection [13]. In addition, if employment rates and childcare attendance rates are already high, the marginal decreases in the costs of childcare would not affect women labor supply anymore [16]. This is also supported by the empirical evidence that the subsidized childcare services are tend to favor women who already enter the workplace with the help of private childcare [17]. That is to say, the subsidies only save some family costs but cannot promote the level of women employment. On account of these mixed results, this article brings up a unique perspective from gender analysis with quantitative evidences from worldwide panel data and Chines data at province level. Our analyses show that the attendance rate of childcare services is positively correlated with the relative employment rate of women to men, which is probably resulted from breaking the vulnerable employment dilemma.

## Global Landscape

We investigate the cross-country education and labor force database in the World Development Indicators (WDIs), the primary World Bank collection on development indicators. Early empirical studies use either the public money invested in the provision of childcare [18] or the density of public childcare services [17] as the indicator. However, more money or more childcare centers does not ensure more a higher ratio of families to choose public childcare services, so that either indicator is indirect. In contrast, this article adopts a direct indicator, namely the enrollment rate $R$ of preprimary school, defined as the ratio of enrolled children to the population with appropriate ages. Another critical parameter under consideration is the female relative employment rate $E$, defined as the female employment rate to the male employment rate for paid work. We analyze the data starting from 1991 when both $R$ and $E$ can be obtained. Since to collect childcare-related data is time-consuming, the missing rate of data after 2018 is larger than 40%. In addition, aggregated records for 41 regional alliances are also eliminated since they are redundant at the presence of country-level data.

After the data cleaning, later reported statistics are resulted from 217 countries and regions, covering a time period from 1991 to 2017. Detailed information and statistics of considered data are shown in the Appendix A and Appendix B, where the former presents the name, code, average values of considered variables over years, and data missing rate of each country or region, and the latter shows average values of considered variables over countries and regions for each year.

We first report the annual Pearson correlation coefficient between the female relative employment rate $E$ and the enrollment rate $R$ of all countries and regions. For each year, a country or region $x$ will be excluded in the calculation if the enrollment rate is less than 1% ($R_x < 1\%$) or the female relative employment rate equals zero ($E_x = 0$). As shown in Fig. 1, in despite of some fluctuations, all correlation coefficients for the 27 years are positive and exhibit an increasing tendency, suggesting significant efficacy of childcare services on women employment worldwide.

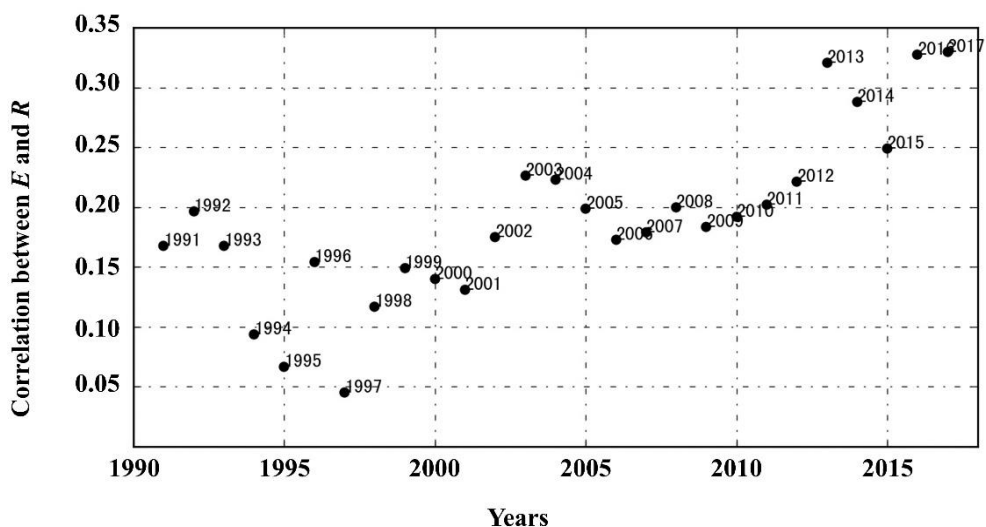

**Figure 1**. The correlations between $E$ and $R$ for years from 1991 to 2017. Each data point denotes the Pearson correlation coefficient of the corresponding year.

Intuitively speaking, personal economic status will affect one's choice on childcare and job. Accordingly, we divide all countries and regions into two categories based on the World Bank standard subject to the per capital of GNI: the high-income group and the non-high-income group (including upper-middle-income, lower-middle-income and low-income ones). The label of each country or region can be found in Appendix A. As shown in Fig. 2, to our surprise, countries and regions in different groups tell different stories. In the high-income group, women are usually of good education background and high income, and thus more likely to attend the market and sending their children to care centers [20]. In addition, governments in high-income group have more resources to truly tackle the gender inequality problems, so that the childcare systems and other related welfare systems are getting more and more thoughtful and humanized. Therefore, we observe higher correlations between $E$ and $R$ and a sharper increasing tendency for high-income group than for all countries and regions (Fig. 2a versus Fig. 1).

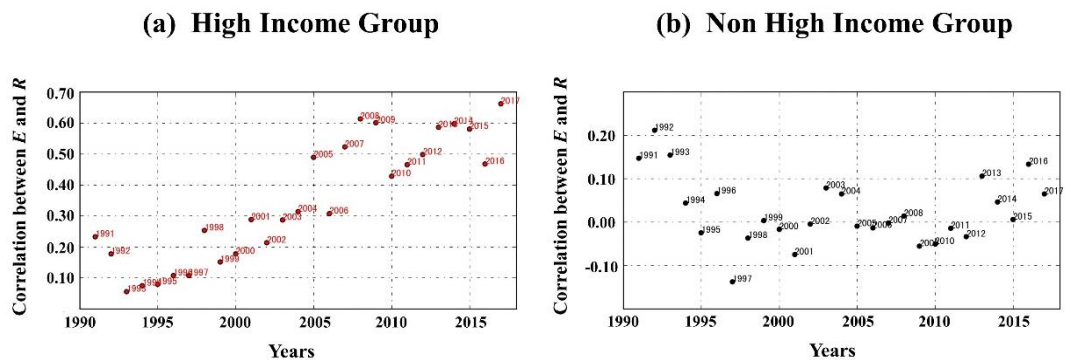

**Figure 2**. The annual correlations between $E$ and $R$ for (a) high-income group and (b) non-high-income group.

In non-high-income group, the cost of childcare centers is a barrier for many families. As the welfare systems in non-high-income group are usually less thoughtful than those in high-income group, the childcare services tend to favor women with formal working positions and regular working schedules, while women assigned with entry-level or lowe-skilled positions usually suffer from unstable schedules including evening and weekend workings. In addition, although the female employment rate is high in non-high-income group, women are often engaged in unpaid subsistence agriculture or family cottage craft, infrequently being involved in the works outside family household [13]. The combination of production and reproduction in one place might lock women into the responsible for parenting, following the traditional motherhood. The above reasons lead to relatively lower enrollment rates of preprimary school in non-high-income group, while the female relative employment rate is not low since women have to make contributions to improve the living conditions. However, as we will analyze later, a considerable fraction of those women is involved in vulnerable employment. In a word, as shown in Fig. 2b, the correlations between $E$ and $R$ are very weak, indicating a potential dilemma in chosing center-based childcare services for women with low income.

# Vulnerable Employment

According to the definition from World Bank, vulnerable employment refers to the workers engaged in unpaid family works and own-account works. Vulnerable workers are less likely to have formal work arrangements, resulted in the shortage of social security, often characterized by unstable working conditions, inadequate earnings and inconsistent working hours. The most critical issue is the inconsistent working hours as the relationship between hours devoting to work and earnings is not linear. The wage and future development of an inconsistent work are remarkably lower and worse than a full-time job with the same position [21]. As a result, transferring women with vulnerable employment to women having formal working positions will largely improve their current income and future development.

The proportion of female vulnerable employment varies greatly among different countries and regions, but is higher than that of male vulnerable employment in general. Here we consider the female relative vulnerable employment rate $V$, defined as the ratio of female vulnerable employment rate to male vulnerable employment rate. From a female perspective, a lower $V$ indicates a better employment condition of women. As shown in Fig. 3, the female relative vulnerable employment rate is strongly negatively correlated with the enrollment rate of preprimary school, suggesting a highly supportive force of childcare services.

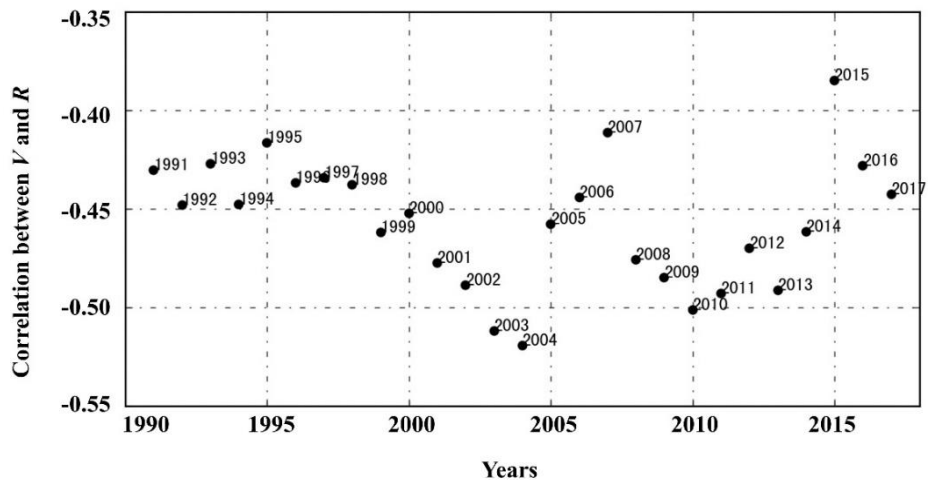

**Figure 3**. The correlations between $V$ and $R$ for years from 1991 to 2017. Each data point denotes the Pearson correlation coefficient of the corresponding year.

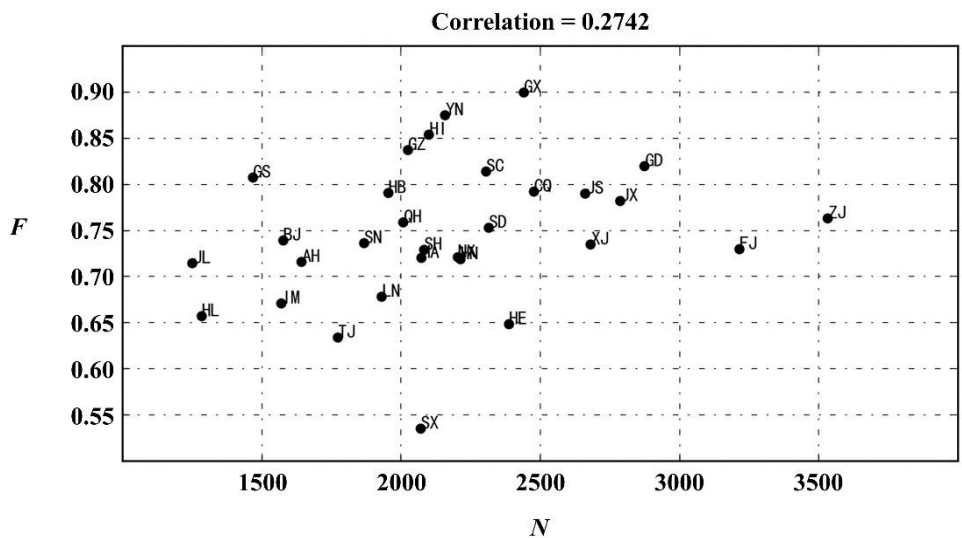

**Figure 4**. The relationship between a province's female relative full-time employment F and average number of kindergarten students per 100,000 population. Each data point corresponds to a province in the year 2010. The average number of kindergarten students per 100,000 population in Tibet is smaller than 1000, which is considered as an outlier and thus excluded in the calculation.

In order to control the institutional and cultural influences, as well as check the robustness of our findings, we further narrow down the study on mainland China. The data is extracted from the National Bureau of Statistic of China, which is different the collection of World Bank. We adopt the female relative full-time employment rate $F$, defined as the ratio of female to male on the proportion of population (aged 16+) working no less than 40 hours per week. The condition of childcare services is quantified by the average number $N$ of kindergarten students per 100,000 population. The names, codes and values of considered parameters of the 31 provinces in mainland China are presented in Appendix C. As shown in Fig. 4, at the province level of mainland China, $F$ is positively correlated with $N$ (the Pearson correlation coefficient is 0.2742), again supporting the above hypothesis that the childcare services can help women in obtaining formal working positions.

# Discussion

In summary, center-based childcare services do have a positive impact on promoting women employment, especially in breaking the vulnerable employment dilemma. However, this influence is affected strongly by the regional economic status, limited the facilitation within the well-developed areas. In the early stage right after the founding of People's Republic of China, the Chinese economics is poor while the Chinese government encourages women to work by setting up a huge number of free childcare centers, including nurseries (mainly for aged 0-3), kindergartens (mainly for aged 3-6) and preprimary schools (one or two years before primary schools). A large fraction of those childcare centers is built and managed by state-owned enterprises. As a result, during that time period, Chinese female labor force participation rate is the highest among all developing countries [22]. Therefore, although the governments in non-high-income group are generally lacking of resources, they should enhance their responsibility to undertake childcare workload [23], not only setting up childcare centers but also making them completely free. Such operation will eventually lead to higher employment rate of women and thus promote the economic development. Moreover, in low-income areas, a considerable fraction of families uses informal childcare services. Therefore, a full picture of the efficacy of childcare services, especially for low-income areas, has to include the existing circumstances and impacts of informal childcare services.

Women's vulnerable employment is a structural challenge. The way to unravel this knot is to have government childcare policy targeting on the vulnerable group as a premise. This needs the intersectional identity perspective and detailed understanding about the requirements and choices of people with difference genders, ages, nationalities, educational levels, socioeconomic status, and so on. Instead of aggregated data used in this article, microdata from large-scale survey is necessary. As we are talking about childcare, a special attention shall be paid on the childcare labor transfer, for the majority workers in the childcare industry are women, and they themselves are the victim of vulnerable employment.

**Acknowledgements**: This work is partially supported by the National Natural Science Foundation of China under No. 11975071.